\documentclass[aps,pra,showpacs,twocolumn]{revtex4}
\usepackage{amsmath}
\usepackage{graphicx}
\usepackage[ansinew]{inputenc}
\usepackage{array}
\usepackage{color}
\usepackage{amsxtra}
\usepackage{amstext}
\usepackage{amssymb}
\usepackage{latexsym}
\begin{document}

\title{Quantum mechanical study of a generic quadratically coupled optomechanical system}

\author{H. Shi and M. Bhattacharya}
\affiliation{School of Physics and Astronomy, Rochester Institute of Technology, 84 Lomb Memorial Drive,
Rochester, NY 14623}

\date{\today}
\begin{abstract}
Typical optomechanical systems involving optical cavities and mechanical oscillators rely on a coupling that
varies linearly with the oscillator displacement. However, recently a coupling varying instead as the square of
the mechanical displacement has been realized, presenting new possibilities for non-demolition measurements and
mechanical squeezing. In this article we present a quantum mechanical study of a generic quadratic-coupling
optomechanical Hamiltonian. First, neglecting dissipation, we provide analytical results for the dressed states,
spectrum, phonon statistics and entanglement. Subsequently, accounting for dissipation, we supply a numerical
treatment using a master equation approach. We expect our results to be of use to optomechanical spectroscopy,
state transfer, wavefunction engineering, and entanglement generation.

\end{abstract}

\maketitle
\section{Introduction}
Optomechanics, which involves the interaction between high-finesse optical resonators and high quality mechanical
oscillators, is currently of great interest as a new frontier in quantum mechanics \cite{Kippenberg2008,Marquardt2009,Schnabel2011}.
Developments that justify such interest include the measurement of mechanical displacements below the standard
quantum limit \cite{Teufel2009,Anetsberger2009}, the observation of back-action effects \cite{Kurn2008}, the optical
cooling of an oscillator to its quantum mechanical ground state \cite{Chan2011} and observation of its zero-point
motion \cite{Amir2012}. These pioneering experiments, as well as many others, have relied on an optomechanical coupling
that varies linearly with the displacement of the mechanical oscillator.

Recently however, an optomechanical coupling that varies \textit{quadratically} with the mechanical displacement has been
experimentally demonstrated in a number of configurations, including membranes \cite{Thompson2008,Jacobs2012} and ultracold
atoms \cite{Purdy2010} in optical cavities, and microdisk resonators \cite{Lin2012}. As opposed to the linearly-coupled
systems, the quadratic coupling offers new possibilities for non-demolition measurements \cite{Thompson2008}, cooling
and squeezing \cite{Jayich2008,Meystre2008,Nunnenkamp2010,Biancafiore2011}, optical springs \cite{Agarwal2008}, and photon
transport \cite{Heinrich2010}.

In this article we present a quantum mechanical study of a generic quadratically coupled optomechanical Hamiltonian that
can model any of the three experimental realizations mentioned above. Our work provides the dressed states of the system, and
clarifies further advantages of quadratic over linear coupling, such as a non-vanishing optomechanical entanglement (in
the absence of dissipation), a higher nonlinearity of the spectrum with respect to photon number, and the possibility of
achieving mechanical squeezing without a projective measurement of the system wavefunction \cite{Mancini1997,Bose1997,Groblacher2009}.
We note that some of the work presented here is complementary to that of Ref.~[\onlinecite{Agarwal2008}], which includes a
description of oscillator occupation probabilities, quadrature squeezing, and back-action effects.

The remainder of this article is arranged as follows. Section ~\ref{sec:Noloss} considers a quadratic optomechanical
system in the absence of noise and dissipation, and presents the eigenstates, eigenvalues, entanglement, phononic properties
and nonclassical state engineering. Section ~\ref{sec:Loss} describes the inclusion of noise and dissipation using a master
equation approach. Section ~\ref{sec:Conc} supplies a conclusion.
\section{Quadratically coupled optomechanical system without noise and dissipation}
\label{sec:Noloss}
The generic Hamiltonian for a quadratically coupled optomechanical system may be written as
\begin{equation}
\label{eq:Hgen}
H=H_{S}+H_{o}+H_{m},
\end{equation}
where $H_{S}$ describes the closed system consisting of an optical mode and a mechanical resonator, while $H_{o}$ and $H_{m}$
describe the coupling of each of these modes to the external environment. The system Hamiltonian is given in detail by
\cite{Cheung2011,Tombesi2011}
\begin{eqnarray}
\label{eq:Ham1}
H_{S}&=&\hbar \omega_{o}a^{\dagger}a+\hbar \omega_{m}b^{\dagger}b
+\hbar g a^{\dagger}a(b^{\dagger}+b)^{2}\nonumber \\
&&+i\hbar E(a^{\dagger}e^{-i\omega_{d}t}-a e^{i\omega_{d}t}),\\
\nonumber
\end{eqnarray}
where $a(a^{\dagger})$ and $b(b^\dagger)$ are the annihilation(creation) operators obeying the bosonic commutation rules
$[a,a^{\dagger}]=[b,b^{\dagger}]=1$, and $\omega_{o}$ and $\omega_{m}$ are the resonant frequencies of the optical and
mechanical modes respectively. The strength
of the optomechanical coupling in Eq.~(\ref{eq:Ham1}) is measured by $g$, whose explicit form depends on the particular
experimental realization \cite{Thompson2008,Purdy2010,Lin2012}. An external field, if present, drives the system at frequency
$\omega_{d}$ and is represented by the coupling
\begin{equation}
\label{eq:EE}
E=\left(\frac{\gamma_{o}P_{in}}{\hbar\omega_{o}}\right)^{1/2},
\end{equation}
where $\gamma_{o}$ is the rate of optical damping and $P_{in}$ is the input power into the optical resonator.

The derivation of Eq.~(\ref{eq:Ham1}) requires several assumptions, as described in detail in Ref.~[\onlinecite{Cheung2011}].
There are two essential approximations. First, it is assumed that photons are not scattered into other modes of the optical
resonator by the mechanical oscillator, i.e. $\omega_{m}$ is much smaller than the free spectral range of the optical cavity.
Second, the amplitude of mechanical motion is assumed to be much smaller than an optical wavelength, which implies that only
the lowest-order dependence of the resonator frequency on oscillator displacement needs to be considered.

In this section, we neglect environmental effects and set $H_{o}=H_{m}=0,$ based on three motivations. First, the physics of
the closed quantum system modeled by the Hamiltonian $H_{S}$ is already quite rich and interesting. Second, it is easier to
understand the system behavior in the absence of external effects. Third, experiment \cite{FreeE2012} and theory \cite{Not2012}
are both moving towards optomechanical configurations in which environmental effects are greatly reduced. If optical loss is
negligible ($H_{o}=0$), it follows that $\gamma_{o}=0$. Also, in the absence of driving, $E = 0$. In that case, Eq.~(\ref{eq:Ham1})
yields \cite{Agarwal2008}
\begin{equation}
\label{eq:Ham2}
H_{Q}=\hbar \omega_{0} a^{\dagger}a+\hbar \omega_{m}b^{\dagger}b
+\hbar g a^{\dagger}a(b^{\dagger}+b)^{2}.
\end{equation}
Clearly the operator $N_{a}=a^{\dagger}a$ commutes with the Hamiltonian of Eq.~(\ref{eq:Ham2}), and thus
the number of photons is conserved. In contrast, $[b^{\dagger}b,H_{Q}] \neq 0$, showing that the number of phonons is not
conserved. We note that zero-drive Hamiltonians have been used by multiple authors to analyze optomechanical systems
quantitatively \cite{Bose1997,Groblacher2009,Rabl2011,Borkje2011}.
\subsection{Eigenstates}
In this section we present the dressed states of the system, the knowledge of which is important to applications such as state
transfer and wavefunction engineering. We assume that each eigenstate of the Hamiltonian of Eq.~(\ref{eq:Ham2}) can be written
simply as a product of the optical state and the mechanical state. Further, since $[a^{\dagger}a,H_{Q}]=0,$ we expect the optical
contribution to the system eigenstate to be simply a number state $|n\rangle_{c},$ where the subscript $c$ stands for `cavity'.
For $g=0$ the mechanical contribution is also a number state $|k\rangle_{m}$. Although this is no longer true if $g \neq 0$,
the form of the third term in Eq.~(\ref{eq:Ham2}) suggests that the optical field `squeezes' the position of the mechanical
oscillator in proportion to the number of optical quanta $n$. Hence we may expect a `squeezed' number state as a contribution
from the mechanical mode. We thus make the ansatz
\begin{equation}
\label{eq:eigen}
|k,n\rangle=S[r(n)]|k\rangle_{m}|n\rangle_{c},
\end{equation}
for the system eigenstate, where
\begin{equation}
\label{eq:squeeze}
S[r(n)]=e^{\frac{r(n)}{2}\left(b^{2}-b^{\dagger 2}\right)},
\end{equation}
is the squeezing operator of quantum optics and $r(n)$ is the cavity photon number dependent dimensionless squeezing parameter.

\begin{figure}
\includegraphics[width=0.46 \textwidth]{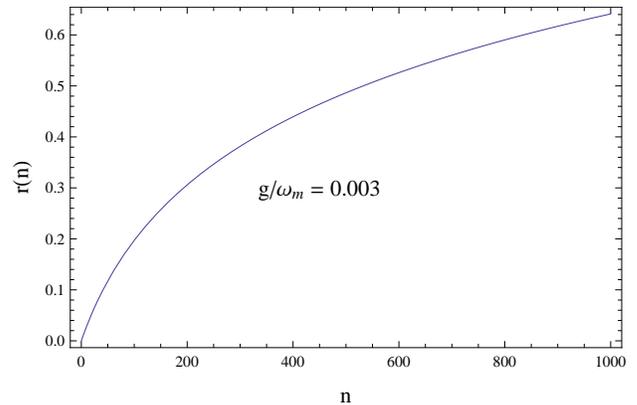}
\caption{The $n-$dependence of the squeezing parameter $r(n)$
of Eq.~(\ref{eq:sf}) for $g/\omega_{m}=0.003$ \cite{Lin2012}.}
\label{fig:QM1}
\end{figure}

We find that $|k,n\rangle$ of Eq.~(\ref{eq:eigen}) satisfies the eigenvalue equation $H|k,n\rangle=E_{k,n}|k,n\rangle$ if $r(n)$
is a positive real number given by
\begin{equation}
\label{eq:sf}
r(n)=\frac{1}{2}\mathrm{\tanh}^{-1}\left[\left(1+\frac{\omega_{m}}{2gn}\right)^{-1}\right],
\end{equation}
which is plotted as a function of the cavity photon number in Fig.~\ref{fig:QM1}
for experimentally relevant parameters \cite{Lin2012}. As in Fig.~\ref{fig:QM1}, since we are primarily interested in the behavior
of the mechanical oscillator, we scale all frequencies and times in our problem appropriately by the mechanical
frequency $\omega_{m}$ in the remainder of the article.

With regard to the eigenstates, we make three observations. First, we find that for $g=0$ we regain from Eq.~(\ref{eq:eigen}) the
expected product $|n\rangle_{c}|k\rangle_{m}$ for the system eigenstate. Second, we note that the dressed states of
Eq.~(\ref{eq:eigen}) are orthonormal, i.e. $\langle k',n'|k,n\rangle=\delta_{k',k}\delta_{n',n},$ unlike the eigenstates provided
in previous studies \cite{Agarwal2008}. Lastly, we warn the reader that $k$ does not denote the number of phonons in an
eigenstate, unless $n=0$.

The eigenstates $|k,n\rangle$ are therefore products of optical number states and squeezed mechanical number states, where the
squeezing parameter for each mechanical state is a function of the number of photons in the corresponding optical number state.
They are analogous to the linear optomechanical coupling eigenstates which are products of optical number states and displaced
mechanical number states, where the displacement parameter for each mechanical state is a function of  the number of photons in
the corresponding optical number state \cite{Groblacher2009,Rabl2011,Borkje2011}.

Since the properties of number states are very well known in quantum optics, we will not describe them here. Squeezed number
states have also been studied earlier in the literature \cite{Kim1989}, and we will only make a comment on the phonon statistics
of the eigenstate. The expectation value of the phonon number is given by
\begin{equation}
\langle k,n|b^{\dagger}b| k,n \rangle=
k +(2k+1)\sinh^{2} r(n),
\end{equation}
which depends on both $k$ as well as $n$, as expected. We also find,
\begin{equation}
\langle k,n |\left(b^{\dagger}b\right)^{2}|k,n\rangle\\
=\langle b^{\dagger}b \rangle^{2}+\frac{(k^{2}+k+1)\sinh^{2}2r(n)}{2}.\\
\end{equation}
Using the above results, we can find analytically the phononic Mandel parameter defined by
\begin{equation}
\label{eq:phonM}
Q_{k}=\frac{\langle (b^{\dagger}b)^{2}\rangle-\langle b^{\dagger}b\rangle^{2}}{\langle b^{\dagger}b\rangle}-1.
\end{equation}
This parameter indicates the nature of the phonon statistics. It is straightforward to verify that an eigenstate $|k,n\rangle$ of Eq.~({\ref{eq:eigen})
can display any type of phonon statistics, depending on the parameters. For example, for $k \neq 0, n=0$ (phononic number state)
we find $Q_{k}=-1$, implying sub-Poissonian statistics. Also, for $k=0, n \neq 0$ (squeezed mechanical vacuum) it follows that
$Q_{k}=\cosh 2 r(n) >1,$ which corresponds to super-Poissonian phonon statistics. Lastly, it can easily be shown that $Q_{k}=0$,
i.e. the eigenstate phonon statistics are Poissonian, if the eigenstate parameters satisfy
\begin{equation}
\sinh r(n)=\left[\frac{\left(4k^{4}+8k^{3}+12k^{2}+8k+1\right)^{1/2}
-\left(2k^{2}+1\right)}{4\left(k^{2}+k+1\right)}\right]^{1/2}.
\end{equation}
\subsection{Eigenvalues}
We now discuss the eigenvalues $E_{k,n}$ of the Hamiltonian of Eq.~(\ref{eq:Ham2}), which are important to applications such as
optomechanical spectroscopy \cite{Groblacher2009} and photon blockade \cite{Rabl2011,Borkje2011} and are given by
\begin{equation}
\label{eq:eigenv}
E_{n,k}=\hbar \left(n\omega_{0}-\frac{\omega_{m}}{2}\right)
+\hbar\left(k+\frac{1}{2}\right)\left[\omega_{m}\left(\omega_{m}+4gn\right)\right]^{1/2}.
\end{equation}
This expression has been derived earlier by Rai et. al \cite{Agarwal2008}. As can be seen in
Fig.~\ref{fig:QM3},
\begin{figure}
\includegraphics[width=0.48 \textwidth]{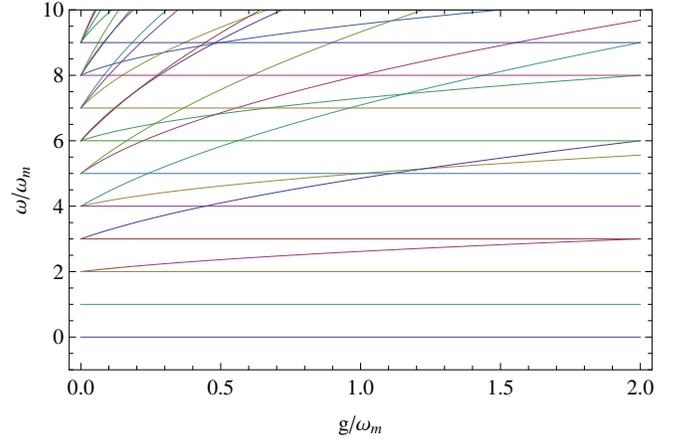}
\caption{The spectrum of the quadratically coupled optomechanical Hamiltonian without driving, see Eq.~(\ref{eq:eigenv}).
For viewing clarity $\omega_{0}=2\omega_{m}$ has been used.}
\label{fig:QM3}
\end{figure}
the spectrum is relatively simple for one-photon excitations, but quite rich when few-photon excitations are considered. It is
interesting to ask how the spectrum is changed by the presence of driving. Therefore, for comparison, we have also shown a
numerical calculation of the Floquet spectrum of Eq.~(\ref{eq:Ham1}) in Fig.~\ref{fig:QM4}, which reveals the effects of driving.
\begin{figure}
\includegraphics[width=0.48 \textwidth]{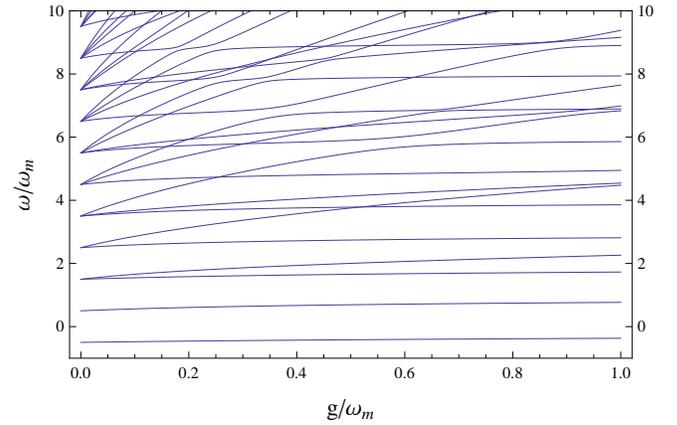}
\caption{The numerically calculated spectrum of the Hamiltonian of Eq.~(\ref{eq:Ham1}) with the same parameters as Fig.~\ref{fig:QM3},
and including driving $(E/\omega_{m}=1)$. It can be seen that some of the crossings of Fig.~\ref{fig:QM3}, namely those between states
differing in photon number by one, have turned into avoided crossings due to the presence of external driving.}
\label{fig:QM4}
\end{figure}

It is worth remarking that the spectrum of the quadratically coupled Hamiltonian is anharmonic to all orders in $n$, unlike the
case of linear coupling, where the anharmonicity is quadratic in $n$ \cite{Groblacher2009,Rabl2011,Borkje2011}. In fact, the anharmonicity
in the present quadratic case is of the same functional form as in the well-known Jaynes-Cummings spectrum, i.e. of the type $(A+Bn)^{1/2},$
where $A$ and $B$ are quantities independent of the photon number $n$. The spectrum described in this section will be observable in the
limit of strong single-photon coupling, which is being approached in optomechanical systems \cite{Rabl2011}.
\subsection{Time evolution of the state vector}
Using the Hamiltonian of Eq.~(\ref{eq:Ham2}) and standard disentangling techniques from quantum optics \cite{Mufti1993},
we can evaluate the time evolution operator in a frame rotating at the optical frequency $\omega_{0}$
\begin{equation}
\label{eq:teo}
U=e^{i\omega_{m}t/2}
e^{\frac{1}{2} \left(\xi^{*}b^{2}-\xi b^{\dagger 2}\right)}
e^{i\eta\left( b^{\dagger}b+\frac{1}{2}\right)},
\end{equation}
where
\begin{equation}
\label{eq:xi}
\xi=e^{i\left(\eta+\frac{\pi}{2}\right)}\sinh^{-1}\left(\frac{r \sin \chi t}{\chi}\right),
\end{equation}
with
\begin{equation}
r=2g a^{\dagger}a,\,\,\chi=\left[\omega_{m}\left(\omega_{m}+4g a^{\dagger}a\right)\right]^{1/2},
\end{equation}
and
\begin{equation}
\label{eq:p}
\eta=\tan^{-1}\left(\frac{p \tan \chi t}{2\chi}\right),
\end{equation}
where
\begin{equation}
\label{eq:PP}
p=-2\left(\omega_{m}+2ga^{\dagger}a\right).
\end{equation}
We note that $\eta$ is purely real.

With the time evolution operator of Eq.~(\ref{eq:teo}) we can, given an initial state vector $|\psi (0)\rangle$, find the vector
$|\psi (t)\rangle=U|\psi (0)\rangle$ at any later time $t$. We will assume as our initial state
\begin{equation}
\label{eq:initw}
|\psi (0)\rangle=|0\rangle_{m}|\alpha\rangle_{c},
\end{equation}
where $|0\rangle_{m}$ is the mechanical ground state, and $|\alpha\rangle_{c}$ is an optical coherent state. Such an initial state
can be realized in experiments which have placed mechanical oscillators in thermal states with occupation number practically equal
to zero \cite{Purdy2010,Chan2011}. Such an initial state has also been used in theoretical analysis of optomechanical systems
earlier \cite{Bose1997}. The state vector at a later time is then found to be
\begin{equation}
\label{eq:wavefn}
|\psi (t)\rangle =\displaystyle\sum_{n=0}^{\infty}c(n,t)|\xi(n)\rangle_{m}|n\rangle_{c},
\end{equation}
where $|\xi(n)\rangle_{m}=S[\xi(n)]|0\rangle_{m}$ (with the $n$-dependence of the squeezing parameter $\xi$ emphasized) is  the squeezed
vacuum state of the mechanical oscillator, and
\begin{equation}
c(n,t)=\frac{\alpha^{n}e^{-\frac{1}{2}\left[|\alpha|^{2}-i\left(\omega_{m}t-\eta \right)\right]}}{\sqrt{n!}}.
\end{equation}
\subsection{Entanglement}
In this section we discuss bipartite entanglement between the optical mode and the mechanical oscillator, which is of interest 
for quantum information processing as well as fundamental studies of quantum mechanics \cite{Bose1997,Ent2007,Rips2013}.

A simple analytical understanding of entanglement in the present system may be arrived at by 
considering Eq.~(\ref{eq:wavefn}). The state $|\psi (t)\rangle$ becomes disentangled when it can be written simply 
as a tensor product of the mechanical and optical states. In order for this to occur either the optical or the mechanical 
state should factor out from the infinite sum of Eq.~(\ref{eq:wavefn}). This can happen only if the mechanical state 
$|\xi(n)\rangle$ in each term is the same. This last condition may be satisfied if $\xi (n)$, the squeezing parameter, 
becomes independent of $n$. From Eqs.~(\ref{eq:xi})-(\ref{eq:p}), it can be seen that the last factor in $\xi$, i.e. 
$\sinh^{-1}(r\sin \chi t/\chi)$, is independent of $n$ only at $t=0$, at which time $\xi (n)\equiv 0$. Thus, although the 
optical and mechanical modes are initially separable, \textit{they are entangled at all subsequent times}. This behavior is 
in complete contrast to the case of linear optomechanical coupling, where in a similarly prepared system, the optical and 
mechanical modes disentangle once every mechanical period \cite{Bose1997}. 

In our discussion of entanglement, we have neglected dissipation and noise. In the presence of these influences, the 
optomechanical states do not remain pure and become mixed, and entanglement is expected to weaken and eventually vanish as 
light ultimately escapes from the resonator. The entanglement of mixed states can be characterized by measures such as the 
logarithmic negativity, which can be computed using, for example, the Quantum Langevin approach \cite{Rogers2012}. As this 
treatment is quite involved and not very revealing in the present case, we have not included it here.
\subsection{Phononic properties}
Knowledge of $|\psi (t)\rangle$ from Eq.~(\ref{eq:wavefn}) allows us to investigate the physical behavior of the mechanical oscillator,
such as the average phonon number $\langle b^{\dagger}b\rangle$, whose time dependence is shown in the absence and presence of dissipation
in Figs.~\ref{fig:QM6}(a) and (b) respectively.
\begin{figure}
\includegraphics[width=0.45 \textwidth]{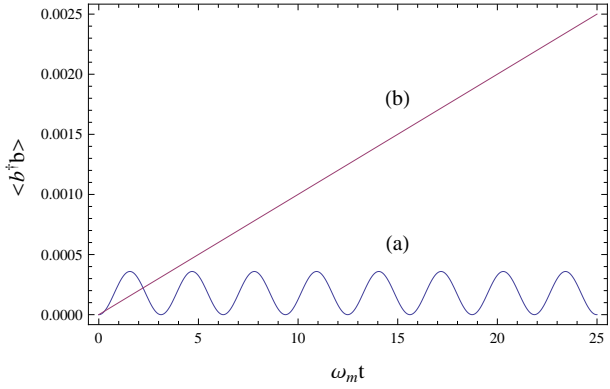}
\caption{Time evolution of the average phonon number $\langle b^{\dagger}b \rangle$ of the mechanical oscillator in the
state Eq.~(\ref{eq:wavefn}). (a) The phonon number evolution in the absence of noise and damping for $g/\omega_{m}=0.003$, and $\alpha=0.1$.
This curve has been scaled up by a factor of a thousand to improve visibility (b) The phonon number evolution obtained from the
master equation of Eq.~(\ref{eq:master}) and including the effects of dissipation ($\gamma_{o}/\omega_{m}=0.1, \gamma_{m}/\omega_{m}=10^{-7},
\bar{n}_{m}=1000, T=100$ mK). It shows the absence of the coherent oscillations and an increase in the oscillator phonon number
upon coming into contact with the thermal reservoir. At longer times, not shown here, $\langle b^{\dagger}b \rangle$ reaches the steady
state value $\langle b^{\dagger}b \rangle = \bar{n}_{m}=1000$.}
\label{fig:QM6}
\end{figure}

We also show the temporal behavior of the phononic Mandel parameter [Eq.~(\ref{eq:phonM})] in Fig.~\ref{fig:QM7}(a). It can be seen
from this figure that in the absence of dissipation, $Q_{k}=-1$ initially as expected for a vacuum state, and subsequently
$Q_{k} \sim 1$ corresponding to a weakly squeezed vacuum state. The oscillations in $Q_{k}$ are due to the periodic time
dependence of $\langle b^{\dagger}b\rangle$ and $\langle \left(b^{\dagger}b\right)^{2}\rangle$. The introduction of dissipation
damps these oscillations as shown in Fig.~\ref{fig:QM7}(b) and at longer times the value of $Q_{k}$ tends to the average number of
bath phonons, as expected for a state in thermal equilibrium.
\begin{figure}
\includegraphics[width=0.45 \textwidth]{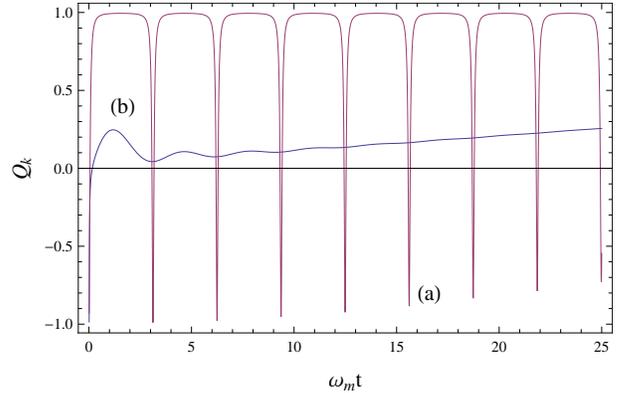}
\caption{The time dependence of the Mandel parameter [$Q_{k}$, defined in Eq.~(\ref{eq:phonM})] for the mechanical oscillator in the
state Eq.~(\ref{eq:wavefn}). (a) The time evolution of the Mandel parameter for $g/\omega_{m}=0.003, \alpha=0.1$ in the absence of
dissipation. From this curve it can be seen that at $t=0$, the Mandel parameter is -1, corresponding to the occupation of
the mechanical ground state. Later in time, $Q_{k}$ reaches the value 1, as expected for a weakly squeezed vacuum state.
The subsequent oscillations are due to the time variation in $\langle b^{\dagger}b\rangle$ and $\langle (b^{\dagger}b)^{2}\rangle$.
(b) Time evolution of the Mandel parameter obtained from the master equation of Eq.~(\ref{eq:master}) including the effects of dissipation ($\gamma_{o}/\omega_{m}=0.1,
\gamma_{m}/\omega_{m}=10^{-7}, \bar{n}_{m}=1000, T=100$ mK). The curve has been magnified by a factor of a hundred to provide better
visibility, and shows the gradual disappearance of the coherent oscillations and the onset of increasing values of $Q_{k}$. At longer
times, not shown in the plot, the Mandel parameter goes to the limit $Q_{k} \rightarrow \bar{n}_{m}$, as expected for a mechanical
state in thermal equilibrium with the bath.}
\label{fig:QM7}
\end{figure}
\subsection{Nonclassical state preparation}
In this section we discuss the preparation of nonclassical states of mechanical motion, which are of interest to metrology
\cite{Bose1997,Borkje2011,Bhattacharya2010}. We focus on mechanical states with sub-vacuum uncertainty (squeezing) in some
quadrature. In general, this quadrature isa  linear combination of position and momentum of the mechanical oscillator.
We find that such squeezed states can be obtained from our system in two ways.

The first technique can be understood from an inspection of the reduced density matrix of the mechanical oscillator,
\begin{equation}
\label{eq:densm}
\rho_{m} = e^{-|\alpha|^{2}}\displaystyle \sum_{n}\frac{|\alpha|^{2n}}{n!}|\xi(n)\rangle\langle \xi^{*}(n)|,
\end{equation}
which shows that the oscillator generally exists in a superposition of squeezed vacua. For a small number of average photons in
the cavity $(\alpha^{2} \sim 1)$, only the first few terms in Eq.~(\ref{eq:densm}) contribute to the quantum state. This can be
verified by examining the time-evolution of the Wigner function corresponding to $\rho_{m}$, as shown in Fig.~\ref{fig:QM8}. It
can be seen that starting initially from the vacuum, the oscillator evolves generally into a state with sub-vacuum uncertainties
in some quadrature. For the largest available experimental coupling, ($g/\omega_{m} \sim 0.003$ \cite{Lin2012}) we found, even in
the absence of dissipation, very little squeezing. To obtain noticeable squeezing we increased the coupling to $g/\omega_{m} = 0.08$.
In that case we found $\sim 1.8$ dB of squeezing at $\omega_{m}t =0.5$, as shown in Fig.~\ref{fig:QM8}. In the presence of dissipation,
as shown in Fig.~\ref{fig:QM9}, the squeezing degrades a little at $\omega_{m}t =0.5$, and more noticeably at $\omega_{m}t =4$. We
note that unitary evolution in the quadratically coupled system yields superpositions of squeezed states rather than pure squeezed
states. Nonetheless, this result is in complete contrast to the case of linear coupling, where unitary evolution cannot produce
nonclassical mechanical states \cite{Bose1997}.
\begin{figure*}
\includegraphics[width=0.80\textwidth]{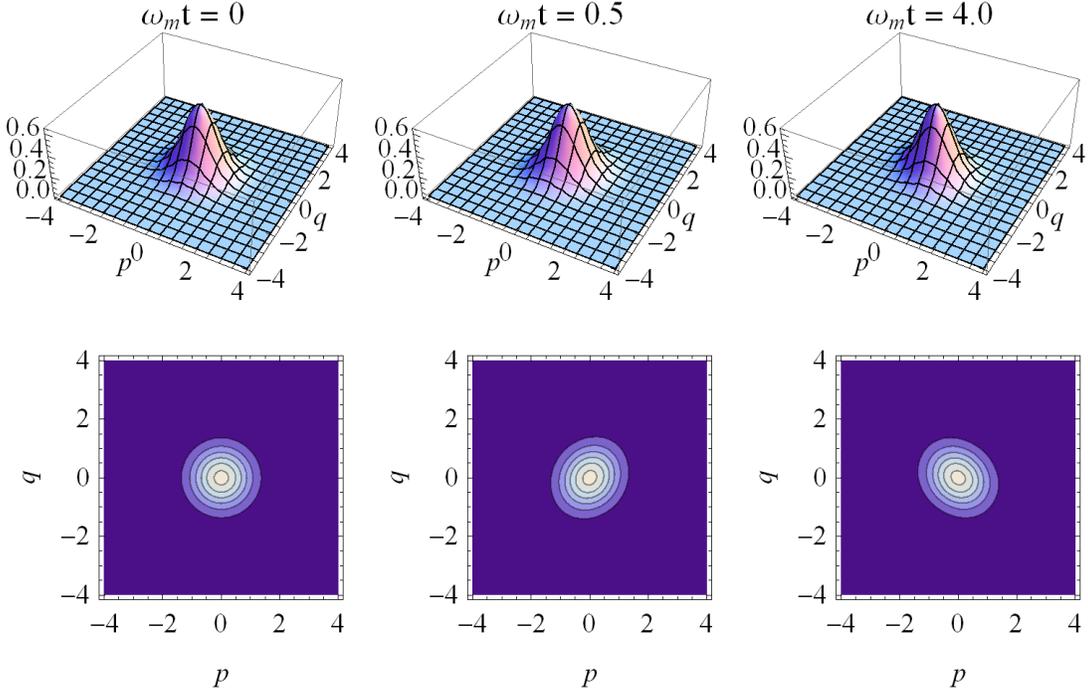}
\caption{Time evolution of the Wigner function of the mechanical oscillator, found from the system wavevector in
Eq.~(\ref{eq:wavefn}); $q$ and $p$ correspond to the position and momentum of the oscillator, respectively. The parameters used are
$g/\omega_{m}=0.08,\alpha = 1$. No dissipation has been included. At $\omega_{m}t=0$ the initially occupied mechanical
vacuum state can be seen. The squeezing at $\omega_{m}t=0.5$ is about $1.8$ dB.}
\label{fig:QM8}
\end{figure*}
\begin{figure*}
\includegraphics[width=0.80\textwidth]{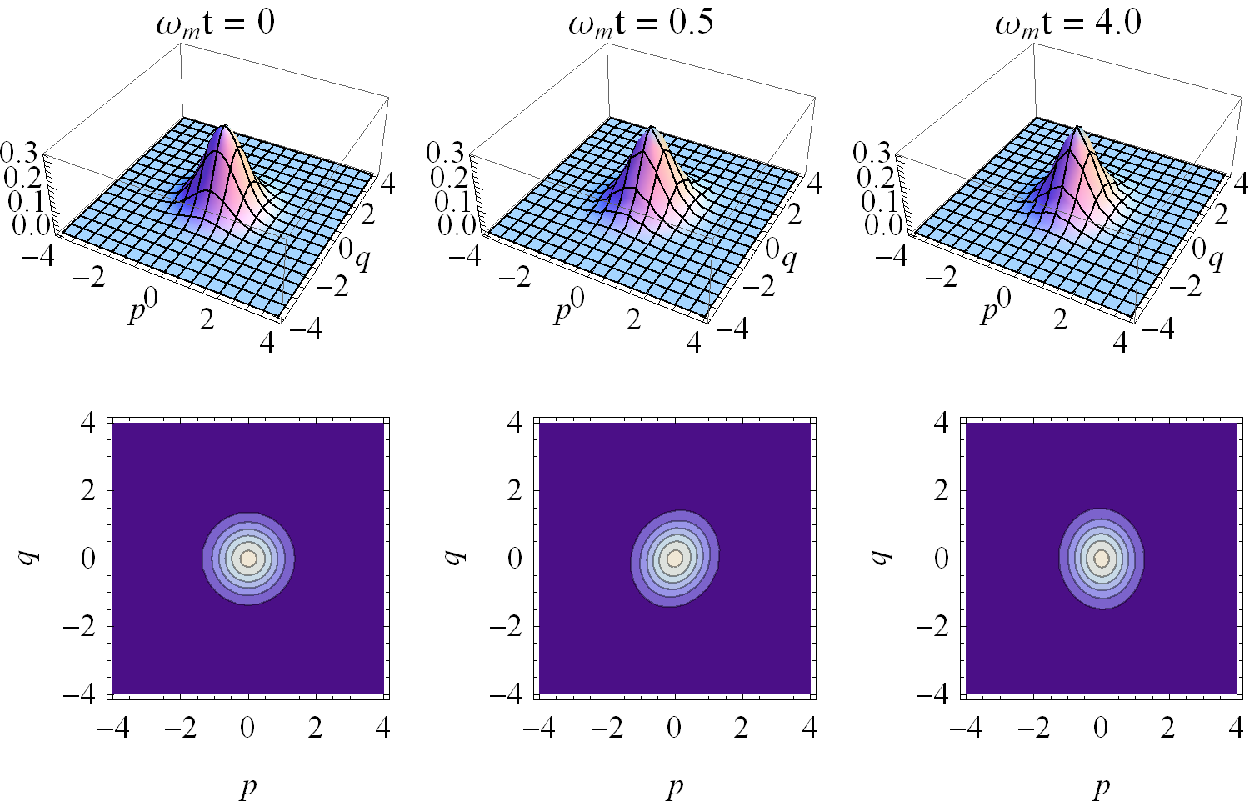}
\caption{Time evolution of the Wigner function of the mechanical oscillator, obtained from the master equation
[Eq.~(\ref{eq:master})]. The parameters used are
$g/\omega_{m}=0.08,\alpha = 1, \gamma_{o}/\omega_{m}=0.1, \gamma_{m}/\omega_{m}=10^{-7}, \bar{n}_{m}=1000, T=100$mK.
Comparing with Fig.~\ref{fig:QM8} it can be seen that dissipation slightly changes the squeezing magnitude and
angle at $\omega_{m}t=0.5,$ and more noticeably at $\omega_{m}t=4$.}
\label{fig:QM9}
\end{figure*}

A more involved method can be used to produce pure squeezed states with quite large squeezing. In this approach a larger
number of average photons in the cavity $(\alpha^{2} \sim 25)$ is required, and in that case several terms in the sum of
Eq.~(\ref{eq:densm}) make a non-negligible contribution to the reduced mechanical density matrix. The corresponding Wigner
function is plotted in Fig.~\ref{fig:QM10},
\begin{figure*}
\includegraphics[width=0.80\textwidth]{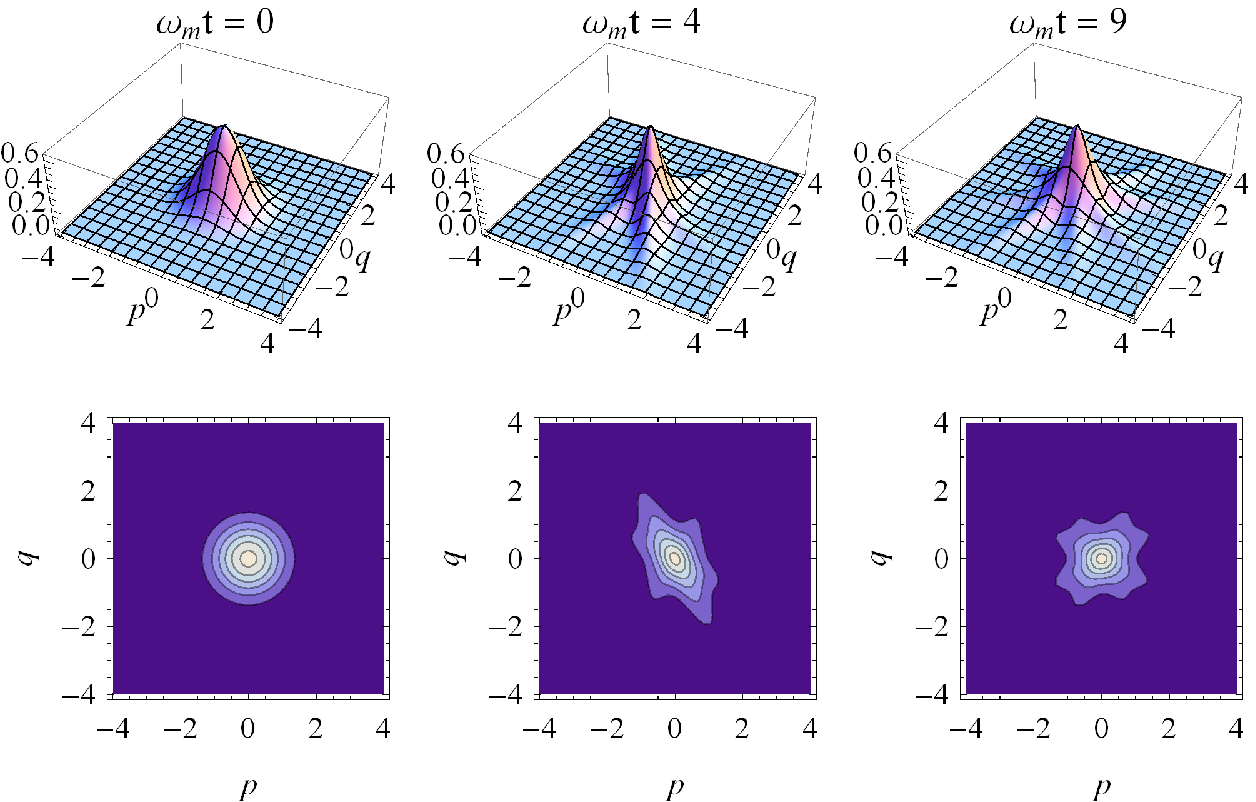}
\caption{Time evolution of the Wigner function of the mechanical oscillator. The parameters used are
$g/\omega_{m}=0.08,\alpha = 5$. No dissipation has been included. At $t=0$ the initially occupied vacuum can be seen.
At $\omega_{m}t=0.5,$ three ellipses, each corresponding to a squeezed mechanical vacuum appear. At $\omega_{m}t=9$, four
ellipses can be detected.}
\label{fig:QM10}
\end{figure*}
and shows multiple mechanical vacua differing in both the magnitude as well as the angle of their squeezing. The technique for
producing pure squeezed states from this superposition follows from the form of the system state vector provided in Eq.~(\ref{eq:wavefn}),
which shows that a conditional measurement of the photon number $n$ in the cavity collapses the mechanical state into the
corresponding squeezed vacuum, $|\xi(n)\rangle_{m}$ \cite{Bose1997}. This procedure can yield quite large mechanical squeezing
of about 10 dB.

The large photon numbers in the cavity required for this technique, and the large phonon numbers resulting from mechanical
squeezing unfortunately make the inclusion of dissipation effects computationally intensive. A treatment of the high photon
number limit for our model as well as an extension to the regime of strong optomechanical coupling will be the subject of a
later work. We note that an analysis of mechanical squeezing at high photon numbers, using a different approach, has been
presented in Ref.[\onlinecite{Nunnenkamp2010}].
\section{Master equation treatment of dissipation}
\label{sec:Loss}
In this section we describe how we account for the presence of environmental effects represented by $H_{o}$ and $H_{m}$ in Eq.~(\ref{eq:Hgen}).
These terms can be handled in a prescriptive manner using the standard master equation approach \cite{CarmichaelBook}. The
master equation is given by
\begin{eqnarray}
\label{eq:master}
\dot{\rho}&=&-\frac{i}{\hbar}[\rho,H_{Q}]+\frac{\gamma_{o}}{2}\left(2 a\rho a^{\dagger}-a^{\dagger}a\rho-\rho a^{\dagger}a\right) \\ \nonumber
&&+\frac{\gamma_{m}}{2}\left(\bar{n}_{m}+1\right)\left(2 b\rho b^{\dagger}-b^{\dagger}b\rho-\rho b^{\dagger}b\right)\\
&&+\frac{\gamma_{m}}{2}\bar{n}_{m}\left(2 b^{\dagger}\rho b-b b^{\dagger}\rho-\rho b b^{\dagger}\right),\nonumber \\
\nonumber
\end{eqnarray}
where $\gamma_{o}$ and $\gamma_{m}$ are the decay rates of optical and mechanical quanta to the respective reservoirs, and
\begin{equation}
\bar{n}_{m}=\frac{1}{e^{\hbar \omega_{m}/k_{B}T}-1},
\end{equation}
is the number of equilibrium quanta of energy $\hbar \omega_{m}$ in the mechanical bath which is held at temperature $T$.

The master equation provided above is valid if $g$ is smaller than either $\gamma_{0}$ or $\gamma_{m}$. This condition ensures
that the regime of strong optomechanical coupling which corresponds to $g>(\gamma_{0},\gamma_{m})$, is avoided. This is because
in the strong coupling regime a master equation has to be derived in the dressed state basis in order to achieve consistency with
thermodynamics \cite{Zoubi2000}. We have also assumed that
the electromagnetic vacuum is effectively at zero temperature. This is a reasonable approximation since there are few
thermal photons available at optical wavelengths
\cite{Mancini1997}. Unless explicitly mentioned, our calculations use experimentally accessible parameters,
$g/\omega_{m}=0.003, \alpha \sim 1, \gamma_{o}/\omega_{m}=0.1, \gamma_{m}/\omega_{m}=10^{-7}, \bar{n}_{m}=1000,$ and $T= 100$ mK
\cite{Thompson2008,Lin2012}.

We have solved the master equation numerically in a frame rotating at the frequency $\omega_{0}$ for the initial state given by
Eq.~(\ref{eq:initw}). The transformation to the rotating frame removes $\omega_{0}$ from the dynamics, and thus we do not
provide its value here. We have then obtained the system density matrix $\rho$ as a function of time. All information subsequently
presented about the system has been obtained from the density matrix. The effects of dissipation have been discussed along with the analytical
results for the dissipationless model in Sec.\ref{sec:Noloss}. We have verified that all our analytical results are reproduced
by the master equation simulation in the absence of dissipation. The effects of driving will be discussed in a future publication.
\section{Conclusion}
\label{sec:Conc}
We have explored the basic quantum characteristics of an important class of optomechanical system where the coupling varies as the
square of the mechanical displacement. We have presented the dressed states, spectrum, time evolution, entanglement, phonon statistics and wavefunction engineering relevant to the system. We have shown that in contrast to the case of linear coupling, quadratic interactions lead to
more persistent entanglement, higher spectral nonlinearity, and a method of generating mechanical squeezing by unitary evolution. We have included dissipation in our analysis for the case of weak optomechanical coupling. Our results should be specifically of use to optomechanical spectroscopy, state transfer, wavefunction engineering, and entanglement generation. More generally, our work will be relevant to quantum information processing and sensing possibilities opened up by optomechanical systems in the quantum regime.
\section{Acknowledgements}
We thank Dr. E. Hach III and Dr. S. Preble for a critical reading of the manuscript.

\end{document}